\documentclass[pra,twocolumn,amsmath, superscriptaddress,notitlepage,showpacs]{revtex4-1}
\usepackage{amsmath,amsfonts,amssymb,amstext,amscd,amsthm,dsfont,bbm,hyperref,natbib,braket}
\usepackage[T1]{fontenc}
\usepackage{physics}
\usepackage{color}
\usepackage{soul,xcolor}
\usepackage{gensymb}
\usepackage{subcaption}
\hypersetup{
    colorlinks = true,
    citecolor=violet,
    linkcolor = red,
    anchorcolor = red,
    citecolor = violet,
    filecolor = red,
    pagecolor = red,
    urlcolor = violet,
    }
    
\usepackage[normalem]{ulem}
\usepackage{graphicx}
\usepackage{bbold}
\usepackage{braket}
\usepackage{placeins}
\usepackage{float}

\usepackage{ulem}

\begin{document}
\title{Experimental realization of quantum non-Markovianity through the convex mixing of Pauli semigroups on an NMR quantum processor}

\author{Vaishali Gulati}
\email{vaishali@iisermohali.ac.in}
\affiliation{Department of Physical Sciences, Indian Institute of Science Education \& Research Mohali,
Sector 81 SAS Nagar, Manauli PO 140306 Punjab India}

\author{Vinayak Jagadish}
\email{vinayak.jagadish@helsinki.fi}
\affiliation{Instytut Fizyki Teoretycznej, Uniwersytet Jagiello{\'n}ski, {\L}ojasiewicza 11, 30-348 Krak\'ow, Poland} 
\affiliation{QTF Centre of Excellence, Department of Physics, University of Helsinki, P.O. Box 43, FI-00014 Helsinki, Finland}

\author{R. Srikanth}
\email{srik@poornaprajna.org}
\affiliation{Theoretical Sciences Division,
Poornaprajna Institute of Scientific Research (PPISR), 
Bidalur post, Devanahalli, Bengaluru 562164, India}

\author{Kavita Dorai}
\email{kavita@iisermohali.ac.in}
\affiliation{Department of Physical Sciences, Indian Institute of Science Education \& Research Mohali,
Sector 81 SAS Nagar, Manauli PO 140306 Punjab India}

\begin{abstract} 
This experimental study aims to investigate the convex combinations of Pauli semigroups with arbitrary mixing
parameters to determine whether the resulting dynamical map exhibits Markovian or non-Markovian behavior. Specifically, we consider the cases of equal as well as unequal mixing of two Pauli semigroups, and demonstrate that the resulting map is always non-Markovian. Additionally, we study three
cases of three-way mixing of the three Pauli semigroups and determine the
Markovianity or non-Markovianity of the resulting maps by experimentally
determining the decay rates.
To simulate the non-unitary dynamics of a single qubit system with different mixing combinations of Pauli semigroups on an NMR quantum processor, we use an algorithm  involving two ancillary qubits. The experimental results align with the theoretical predictions.
\end{abstract}

\maketitle  
\section{Introduction}
The field of quantum computing is rapidly developing, and there is a crucial need to develop reliable methods to characterize and control quantum systems. Quantum systems can interact with their environment in various ways, leading to
decoherence and dissipation, which could have a deleterious effect on the computational protocols. The study of open quantum systems~\cite{petruccione,haroche_exploring_2006} therefore has significant
implications for applications in quantum information processing, quantum
computing, and quantum communication. Recent research has focused on the effect of decoherence on the performance of quantum computers~\cite{Knill-nature-2005} and the use of error correction codes to
address this issue~\cite{fowler-pra-2012}. A critical aspect of open quantum systems is characterizing their dynamical behavior, with a particular focus on the distinction between Markovian and non-Markovian dynamics~\cite{breuer_colloquium:_2016,li_concepts_2017,de_vega_dynamics_2017}. The theory of
non-Markovian dynamics has become an important area of research, with a focus on characterization, quantification, and detection of non-Markovian behavior~\cite{carmele-nano-2019,zhang-ieee-2019,jiang-pra-2013}.

The reduced dynamics of the quantum system of interest undergoing open evolution is described by a time-continuous family of completely positive (CP) and trace-preserving (TP)  linear maps $\{\Lambda (t): t\geq 0, \Lambda(0) = \mathbbm{1}\}$ known as the quantum dynamical map, acting on the bounded operators of the Hilbert space of the system of interest~\cite{sudarshan_stochastic_1961, Quanta77}.
The dynamical map is also related to the time-local generator $\mathcal{L}(t)$~\cite{gorini_completely_1976} in the time-local master equation, $\dot{\Lambda}(t) = \mathcal{L}(t)\Lambda(t)$, with 
\begin{equation}
\begin{split}
 \mathcal{L}(t)[\rho]=&-\imath [H(t),\rho]\\&
+\sum_i \gamma_i (t)
\left(L_i(t)\rho L_i(t)^\dagger-\frac {1}{2}\{L_i(t)^\dagger L_i(t),\rho\}\right),
\label{gksl}
\end{split}
\end{equation}
were $H(t)$ is the effective Hamiltonian, $L_i(t)$'s are the noise operators, and $\gamma_i (t)$ the decoherence rates. The divisibility of the dynamical map is expressed as follows.
\begin{equation}
\Lambda(t_f, t_i) = V(t_f, t)\Lambda(t, t_i),\quad \forall t_f\geq t \geq t_i\geq 0.
\label{cpdivdef}
\end{equation}
The map is CP-divisible if for all $t$, the propagator $V(t_f, t)$ is CP and the corresponding decay rates $\gamma_i (t)$ are positive at all times.  Otherwise, the map is said to be CP indivisible.

In contrast with classical non-Markovianity, quantum non-Markovianity does not have a unique definition~\cite{breuer_colloquium:_2016,li_concepts_2017,rivasreview}. Two major proposals to address quantum non-Markovianity,  are based  on the CP-indivisibility criterion (RHP)~\cite{rivas_entanglement_2010, hall2010}  and on the distinguishability  of  states (BLP)~\cite{breuer-prl-2009,laine-pra-2010}. According to the RHP divisibility criterion~\cite{rivas_entanglement_2010}, a quantum dynamical map is non-Markovian if it is CP-indivisible. A Markovian evolution, therefore is CP-divisible, with all the decay rates $\gamma_{i}(t)$ in the time-local master equation Eq. (\ref{gksl}) are positive at all times. A temporarily negative decay rate is therefore a signature of CP-indivisibility of the map and therefore non-Markovianity.  According to the BLP definition~\cite{breuer-prl-2009}, a quantum dynamical map $\Lambda(t)$ is said to be Markovian if it does not increase the distinguishability of two initial states $\rho_1$ and $\rho_2$, i.e., if $\Vert\Lambda(t)(\rho_1) - \Lambda(t)(\rho_2)\Vert \le \Vert\Lambda(0)(\rho_1) - \Lambda(0)(\rho_2)\vert\vert$, where $\Vert \cdot \Vert$ denotes the trace distance. In this work, we stick to the CP-indivisibility criterion of non-Markovianity.

Convex combinations of Pauli semigroups and time-dependent Markovian Pauli dynamical maps was studied in~\cite{jagadish_convex_2020,jagadish_nonqds_2020} discussing the geometrical aspects and non-Markovianity. These results showed the non-convexity of the sets of CP-divisible and CP-indivisible Pauli dynamical maps. Convex combination of semigroups of generalized Pauli dynamical maps has been addressed in~\cite{siudzinskajpa2020}. In~\cite{megier_eternal_2017}, it was shown that an eternally non-Markovian evolution  arises from a mixture of Markovian semigroups. 
Convex combinations of noninvertible dynamical maps has also been studied recently~\cite{siudzinska_markovian_2021,jagadish2022noninvertibility,jagadish_measureinvert_2022,jagadish_nonivertible_2023}. For the case of generalized Pauli dynamical maps, it was shown that mixing invertible maps can never result in noninvertible maps~\cite{siudzinska_markovian_2021}. Subsequently, it was also shown that noninvertibility of the generalized Pauli input maps is necessary for getting a semigroup~\cite{jagadish2022noninvertibility}. The fraction of (non)invertible maps obtained by mixing noninvertible generalized Pauli maps was quantified in~\cite{jagadish_measureinvert_2022}. The measure of the set of non-Markovian maps obtained by mixing noninvertible Pauli maps was studied in~\cite{jagadish_nonivertible_2023}.

In recent years, there has been a growing interest in the experimental
implementation of non-Markovian dynamics in various physical systems, including quantum dots ~\cite{liu-prl-2019,harouni-cpb-2020,fux-prappl-2021}, superconducting qubits
~\cite{zhang-prappl-2022}, trapped ions~\cite {li-prl-2022,li-epl-2020}, and nuclear magnetic resonance (NMR) systems~\cite{matsuzaki-njp-2019,bengs-jmr-2021}. NMR systems, in particular, are a useful platform to investigate non-Markovian dynamics due to their excellent
ability to control and manipulate system-environment interactions. Various studies in NMR investigate different quantum correlations present in the system~\cite{vg-epjd-2022,singh-pra-2018} and their dynamics under various
environments~\cite{singh-pra-2018-3udd,akan-qip-2022}.

In this work, we aim to experimentally study the behavior of a single qubit system under the effect of different mixing combinations of Pauli semigroups on an NMR quantum processor. We demonstrate that the mixing of any two Markovian Pauli semigroups produces a map which is CP-indivisible and therefore RHP non-Markovian. One of the decay rate always turns out to be negative in this scenario. We also verify our experimental results for arbitrary choices of the mixing
parameters for the dynamical semigroup realizations of the three Pauli semigroups which are in agreement with the notion of Pauli Simplex as defined in ~\cite{jagadish_convex_2020}. We note that the non-Markovian nature of the map becomes apparent when one or more of the decay rates becomes negative.  We consider the case of a single qubit with two ancilla qubits to
simulate non-unitary dynamics and make use of the algorithm for the
circuit design as in~\cite{xin-pra-2017}.

The rest of this paper is organized as follows. Sec.~\ref{paulisec} briefly describes the theory of the convex combinations of Pauli semigroups. The experimental details and results are presented in Sec.~\ref{exp}. We then conclude in Sec.~\ref{concl}.

\section{Convex Combination of Pauli semigroups}
\label{paulisec}
Consider the three Pauli dynamical semigroups, 
\begin{eqnarray}
\label{paulichanndef}
\Lambda_i (t)[\rho]&=&[1-p(t)]\rho + p(t)\sigma_i\rho \sigma_i, \thinspace i= 1,2,3, \mathrm{with}\nonumber\\
p(t) &=& \frac{1-e^{-ct}}{2},\thinspace c >0.
\end{eqnarray} 
Here $p(t)$ is the decoherence function and $\sigma_i$ are the Pauli matrices.   

The convex combination of the three Pauli semigroups Eq.~(\ref{paulichanndef}), each mixed in proportions of $x_i$ is,
\begin{equation}
\label{outputmappauli}
\tilde{\Lambda}(t) = \sum_{i=1}^{3} x_{i} \Lambda_i (t),  \quad (x_i >0, \sum_i x_i =1).
\end{equation}
Let us call the three $\Lambda_i (t)$'s input maps and $\tilde{\Lambda}(t)$ the output map. The associated time-local master equation for $\tilde{\Lambda}(t)$ is
\begin{equation}
\label{megen}
\mathcal{L}(t)[\rho] = \sum_{i=1}^{3}\gamma_{i} (t) (\sigma_i\rho\sigma_i-\rho),
\end{equation}
with the decay rates
\begin{widetext}
\begin{eqnarray}
\gamma_1(t) &=& \left(\frac{1-x_2}{1-2 (1-x_2)p(t)}+\frac{1-x_3}{1-2 (1-x_3)p(t)}-\frac{1-x_1}{1-2 (1-x_1) p(t)}\right)\frac{\dot{p}(t)}{2}\nonumber\\
\gamma_2(t) &=&\left(\frac{1-x_1}{1-2 (1-x_1)p(t)}+\frac{1-x_3}{1-2 (1-x_3)p(t)}-\frac{1-x_2}{1-2 (1-x_2) p(t)}\right)\frac{\dot{p}(t)}{2}\nonumber\\
\gamma_3(t) &=&\left(\frac{1-x_1}{1-2 (1-x_1)p(t)}+\frac{1-x_2}{1-2 (1-x_2)p(t)}-\frac{1-x_3}{1-2 (1-x_3) p(t)}\right)\frac{\dot{p}(t)}{2}. 
\label{ratesdecaythree}
\end{eqnarray}
\end{widetext}
The CP-divisibility and therefore, the Markovianity of output map $\tilde{\Lambda}(t)$ depends on the mixing coefficients $x_i$. For instance, an equal mixing of the three Pauli semigroups results in a Markovian output. The fraction of non-Markovian (CP-indivisible) maps obtained by mixing Pauli semigroups was reported in~\cite{jagadish_convex_2020}.
As opposed to three-way mixing, any mixing of two Pauli semigroups is always non-Markovian. To this end, let $x_1=0$. The decay rate, $\gamma_1(t)$ turns out to be
\begin{equation}
\gamma_1(t) = -\left[\frac{ (1-x_2) x_2 [1-p(t)] p(t)}{[1-2p(t)] [1-2 (1-x_2) p(t)] [1-2 x_2 p(t)]}\right]  \dot{p}(t),
\label{ratesdecay}
\end{equation}
which remains negative for all values of $x_2$. (Note that $x_3 = 1-x_2$.)

\section{Experimental Analysis of Markovianity and non-Markovianity}
\label{exp}
\subsection{NMR Simulation of Pauli semigroups}
\label{results}
 A dynamical map acting on a system of $d$-dimensional Hilbert space could be simulated by a $d^2$-dimensional ancilla if one allows the most general unitary evolution
of the total system under the assumption that the ancillae is initialized in a pure state~\cite{schumacher96}. Therefore, to simulate maps on a qubit, a two qubit ancillae 
is sufficient.
The finite time map $\Tilde{\Lambda}(t)$,as in Eq.~(\ref{outputmappauli}) being CPTP admits an operator-sum representation, $\tilde{\Lambda}(t)(\rho) = \sum_k E_k(t)\rho E^\dag_k(t)$, where the operators
$E_k(t)$ satisfies the trace-preservation condition, $\sum_kE^\dag_k(t)E_k(t)=\mathbb{1}$.  

The non-unitary operators $E_k (t)$ associated with the dynamical map can be decomposed into a linear combination of 4 unitary operators  (Pauli matrices $\sigma_i$'s in this case) and are experimentally implemented using 2 ancillary qubits added to the working system.   Efficient implementation of the non-unitary transformation represented by $\Tilde{\Lambda}(t)$ is achievable when suitable unitary operations $U, V$, and $W$ are found, such that $E_k = \sum_i W_{ki}V_{i0}U_i$. 
By applying the overall unitary operation $(I\otimes W)U(I\otimes V)$ to the initial state of the working system and ancillary system, followed by the trace-out of the ancilla, the simulation of the map is obtained. The algorithm involving three unitaries offers the advantage in implementing the maps involving the convex mixtures of Pauli semigroups in a more general manner. This approach eliminates the need to design separate circuits for each specific mixing combination. By incorporating three unitaries into the algorithm, it becomes possible to dynamically adjust and experiment with different mixing parameters and Pauli operators, allowing for greater flexibility and versatility in simulating the desired non-unitary dynamics.
The algorithm is as follows.

\begin{itemize}
\item Transforming the  state of the ancilla qubits: After initializing the three-qubit system in the state $|0\rangle_s|00\rangle$ where
$|0\rangle_s$ is the state of the system qubit  and $|00\rangle$ that of the ancillary qubits, a unitary operation $V$ is performed on the ancillary qubits. The composite state evolves to $V_{00}|0\rangle_s|00\rangle + V_{10}|0\rangle_s|01\rangle + V_{20}|0\rangle_s|10\rangle+ V_{30}|0\rangle_s|11\rangle$. The mixing parameters and the decoherence function associated with the Kraus operators determine the values in the first column of the unitary matrix $V$.

\item Transforming the  state of the system: The unitary operations $\sigma_i$ are applied
on the system qubit depending on the state of the ancilla qubits acting as control qubits. 
\begin{equation}
\label{cu}
    U = \sigma_0
\otimes|00\rangle\langle00|+\sigma_1 \otimes|01\rangle\langle01|+\sigma_2 \otimes|10\rangle\langle10|+\sigma_3 \otimes|11\rangle\langle11|, 
\end{equation} 
where $\sigma_0$ is the Identity matrix. The system now evolves to the state $V_{00}\sigma_0|0\rangle_s|00\rangle + V_{10}\sigma_1|0\rangle_s|01\rangle + V_{20}\sigma_2|0\rangle_s|10\rangle+ V_{30}\sigma_3|0\rangle_s|11\rangle$.

\item Finally, 
the unitary
operation $W$ is performed on the ancillary system which transforms the state into 
$\sum_{i,k=0}^{3} W_{ki}V_{i0}\sigma_i
|0\rangle_s |k\rangle$, where $E_k
=\sum_{i=0}^{3} W_{ki}V_{i0}\sigma_i$. The elements of matrix $W$ are uniquely determined by the choice of matrix elements of $V$. We obtain the $W$ matrix as Identity matrix in our cases.

\item On measuring the final state of the working system with the ancillary system in the state $|k\rangle\langle k|$, we obtain $E_k|0\rangle_s\langle0|_sE^{\dagger}_k$. By tracing out the ancillary qubits, summing over each state $|k\rangle\langle k|$, the resultant is $ \sum_k E_k(t)|0\rangle_s\langle0|_s E^\dag_k(t)$ which corresponds to simulating the map $\tilde{\Lambda}(\rho)$ where the initial state of the system $\rho$ is $|0\rangle\langle0|$.
\end{itemize}

The specific forms of the matrices  $V$ used in the experiments depend on the
dynamical map under consideration, and the specific forms used in our experiments are given in the following section. 
\begin{figure}[h]
\centering
\includegraphics[scale=1]{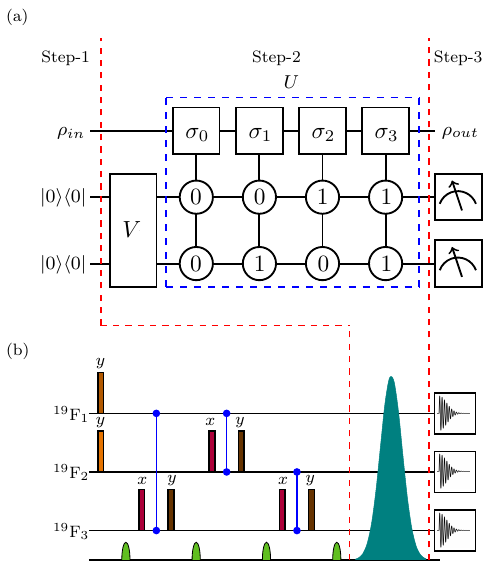}
\caption{(a) Schematic of the circuit used to simulate the dynamical map
obtained from the convex combination of two and three Pauli dynamical maps. For both
two- and three-dynamical map mixing, the controlled operation $U$ is the same.
$\sigma_i$ denote the Pauli matrices, with $\sigma_0$ being Identity matrix. The unitary operation $V$ is
different for the cases of two-way and three-way mixing.  The $W$ operation is equivalent to
the Identity operation and hence not implemented experimentally.  (b) The NMR
pulse sequence used to simulate the map. The rectangular shapes
represent radiofrequency (rf) pulses of differing angles and phases (which are written on the top
of each pulse).  CNOT operations between two qubits are represented by blue
lines between the corresponding qubits.  Step 1 corresponds to the preparation
of the input state.  Gradient pulses are represented by shaped green curves,
while the GRAPE-optimized pulse to implement Step 2 of the  circuit is
represented by a large dark green curve, applied simultaneously on all three
qubits.  Step 3 corresponds to measurements on all the three qubits. }
\label{figure1}
\end{figure}

\subsection{Experimental Parameters}
\label{parameters}
\begin{figure}[h]
\centering
\includegraphics[scale=1]{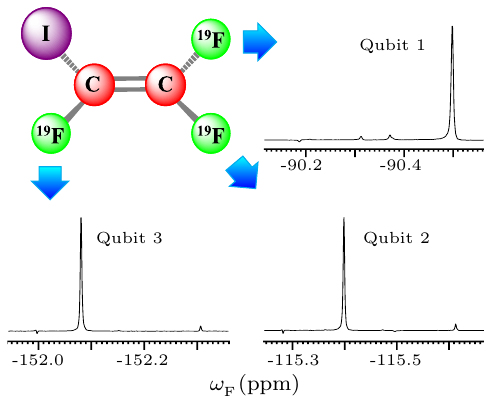}
\caption{The structure of the molecule trifluoroiodoethylene with three NMR active spin$-1/2$ $^{19}$F nuclei acting as three qubits, along with the NMR spectra of the pseudo pure state $\vert 000 \rangle$ which represents the initial state of the three-qubit system. The $x$-axis represents the frequency scale presented in parts per million (ppm) as commonly observed in the standard Bruker spectrometers. The negative values on the $x$-axis represent the frequency offset from the reference frequency indicating upfield shifts.}
\label{figure_mol}
\end{figure}

The three NMR qubits were realized
using the three ${}^{19}$F spin-1/2 nuclei in the
molecule trifluoroiodoethylene (Fig.~\ref{figure_mol}) 
dissolved in the deuterated solvent, d6-acetone. 
All experiments were performed at ambient
temperature ($\approx 298$~K) on a Bruker AVANCE-III 400 MHz NMR spectrometer
equipped with a Broadband Observe (BBO) probe.  The high-temperature, high-field approximation simplifies the NMR Hamiltonian by neglecting certain terms when the thermal and Zeeman energies dominate over other interactions. This approximation enables easier analysis and calculations in NMR experiments. The resulting Hamiltonian, assuming weak scalar coupling $J_{ij}$ between spins $i$ and $j$, is given by~\cite{oliveira-book} 
\begin{equation}
{\cal H} = - \sum_{i=1}^{3} \omega_i I_{iz}
+ 2 \pi \sum_{i<j}^{3} J_{ij} I_{iz} I_{jz},
\end{equation}
where $\omega_i$ is the chemical shift of the $i$th spin, and  $I_{iz}$
represents the $z$-component of the spin-$\frac{1}{2}$ operator for the $i$th
spin.

Nuclear spins at thermal equilibrium are
represented by the density operator,
\begin{equation}
\rho =\frac{\exp(-H/k_{B}T)}{Z},
\end{equation}
where $H$ is the Hamiltonian of the system, $k_{B}$ is 
the Boltzmann's constant, $T$ is the temperature, 
and $Z$ is the partition function.

Starting from thermal equilibrium, the system is prepared in a pseudopure
state (PPS) using the spatial averaging technique~\cite{cory-1998,avikmitra},
with the density matrix corresponding to the PPS being given by
\begin{equation}
\rho_{000} = \frac{(1- \epsilon)}{8}\mathbbm{1}_8 +\epsilon
\vert 000 \rangle \langle 000\vert,
\label{pps-eqn}
\end{equation}
where $\epsilon \sim 10^{-5}$ is the spin polarization at
room temperature and $\mathbbm{1}_8$ is the $8 \times 8$
identity operator. The identity part of the density operator plays no role 
and the NMR signal arises solely
from 
the traceless part of the density matrix given in Eq.~(\ref{pps-eqn}).

$T_1$ and $T_2$ relaxation times in NMR describe the return to equilibrium and loss of phase coherence of nuclear spins. $T_1$ measures the recovery of longitudinal magnetization, while $T_2$ measures the decay of transverse magnetization.
However, the faster decay of transverse magnetization observed in practice is often attributed to $T^{*}_{2}$ relaxation, which combines intrinsic $T_{2}$ relaxation and magnetic field variation effects.$T^{*}_{2}$ for our experimental setup yields a value of approximately 0.1869 $s$.  The experimentally
measured scalar couplings are given by  $J_{12}$= 69.65 Hz, $J_{13}$= 47.67 Hz
and $J_{23}$= -128.32 Hz.

The radiofrequency (rf) required for creating the PPS state were
designed using the Gradient Ascent Pulse Engineering
(GRAPE)~\cite{Khaneja-jmr-2005} technique, along with pulsed 
magnetic field gradients~\cite{shruti-ijqi}.
 The GRAPE pulses obtained are for the collective operation of $U$ and $V$ at each time point. To clarify, for each time point, a specific unitary matrix is obtained by the product of $U$ and $V$. The GRAPE pulse length varies according to different unitaries simulated at different time points.  For instance, at $t$ = 0.1$s$, the GRAPE pulse length is approximately 700 $\mu$s, and at $t = 1.5s$, it is approximately 2500 $\mu$s.
The system was evolved from the PPS to the other states
via state-to-state transfer unitaries, and all states
were created with high fidelities $\geq
0.99$.
The standard methods for quantum state reconstruction for NMR quantum
information processing typically involve performing full state
tomography~\cite{long-qst,leskowitz} which is computationally expensive,
although some alternatives involving maximum likelihood estimation have been
proposed and used~\cite{singh-pla-2016}.  For this work, we 
used a least squares constrained convex optimization method to reconstruct the
density matrix of the desired 
state~\cite{gaikwad-ijqi-2020,akshay-scirep}. 
Fidelities of the
experimentally reconstructed states (as compared to the theoretically expected
state) were computed using the measure~\cite{cory_prl_1998,weinstein_prl_2001},
\begin{equation} 
{\mathcal F}(\chi^{}_{\rm
expt},\chi^{}_{{\rm theo}})= \frac{\vert{\rm Tr}[\chi^{}_{\rm
expt}\chi_{\rm theo}^\dagger]\vert} {\sqrt{{\rm Tr}[\chi_{\rm
expt}^\dagger\chi^{}_{\rm expt}] {\rm Tr}[\chi_{\rm
theo}^\dagger\chi^{}_{\rm theo}]}},
\label{e12}
\end{equation}
where $\chi^{}_{{\rm theo}}$ and $\chi^{}_{{\rm expt}}$ denote the theoretical
and experimental density matrices respectively.  We experimentally prepared
the PPS with a fidelity of
0.9979 $\pm$ 0.0001. The PPS fidelity without convex optimization, calculated with the linear inversion method, is 0.9933 $\pm$ 0.0005.

\subsubsection{Mixing of Two Pauli Semigroups}
We experimentally demonstrate mixing of two-Pauli semigroups for two cases each with the decoherence parameter
$p(t) = [1-\exp{(-2t)}]/2$. To this end, we consider convex mixing as
\begin{equation}
\tilde{\Lambda}(t)(\rho) = a\Lambda_3(t)(\rho)+(1-a)\Lambda_2(t)(\rho).
\end{equation}
The two cases considered are
\begin{itemize}
    \item Equal mixing with the mixing parameter $a=0.5$ and 
\item unequal mixing with the mixing parameter $ a=0.25$.
\end{itemize}
For the simulation of mixing two Pauli semigroups, the algorithm described above leads to the following matrix.
\begin{eqnarray}
V = \left(
\begin{array}{cccc}
 \sqrt{1-p(t)} & \sqrt{p(t)} & 0 & 0 \\
 0 & 0 & 1 & 0 \\
 \sqrt{p(t)(1-a)} & -\sqrt{(1-a) (1-p(t))} & 0 & \sqrt{a} \\
 \sqrt{a p(t)} & -\sqrt{a(1- p(t))} & 0 & -\sqrt{1-a} \\
\end{array}
\right).
\end{eqnarray}

To implement the unitary for the convex combination of the case of mixing two and three Pauli semigroups experimentally, we
utilized the quantum circuit shown in Fig.~\ref{figure1}. For mixing of both two and three semigroups, the controlled operation $U$ is the same, as in Eq.~\ref{cu}. The unitary operation $V$ is
different for the two-way and three-way mixing. The $W$ operation is equivalent to
the Identity operation for both cases and is hence not implemented experimentally. For the implementation of the NMR
pulse sequence, GRAPE-optimized pulses are used. The unitaries $U$ and $V$ are designed so as to be implemented by use of a single pulse for each time point in all the cases. 
The experimental procedure involves three steps. 
\begin{itemize}
\item Step 1- Initialization: The system is prepared in the state $|000\rangle\langle000|$ 
with the help of optimized pulses and magnetic field gradients.
\item Step 2- Simulation of the non-unitary dynamics: The implementation of $U$ and $V$ with GRAPE optimized pulses.
\item Step 3- Measurement: The acquisition and tomography pulses are applied.
\end{itemize}

 The rectangular shapes in Fig.~\ref{figure1} depict the rf pulses used to prepare the initial pseudopure state required for step 1 of the algorithm. Each rectangle is associated with specific phases, which are indicated above them. The magnetic field direction is assumed to align with the $z$-axis. The rf pulses are applied along the $x$ or $y$-axis at specific angles, allowing precise control over qubit rotations and transformations.
With the knowledge of the desired phases and angles of the rf pulses, we can perform operations like single-qubit rotations and two-qubit gates. For example, the first qubit is rotated by an angle of $\theta_1 =\frac{5\pi}{12}$ radians around the $y$-axis, while the second qubit is rotated by an angle of $\theta_2 =\frac{\pi}{3}$ radians. CNOT operations
between two qubits are represented by blue lines
between the corresponding qubits. The complete pulse sequence corresponding to the CNOT gate can be found in~\cite{vg-epjd-2022}. Before the CNOT gate operation, an $x$ pulse with an angle of $\frac{\pi}{4}$ is applied. This pulse rotates the state of the qubit around the $x$-axis. Following the CNOT gate, a $y$ pulse with an angle of $-\frac{\pi}{4}$ is applied, which rotates the state around the $y$-axis. The angles and pulses of the RF pulses or gate operations are carefully chosen to achieve the desired output state or perform the targeted operation. The specific choice of angles or gates depend on our goal which in this case is to prepare the PPS.
After the initialization, a GRAPE pulse corresponding to Step 2 of the algorithm is applied. This pulse applies the unitary operations $V$ and $U$, depending on the specific case being considered.

\subsubsection{Mixing of Three Pauli Semigroups}
We next consider the case of the convex combination of three Pauli semigroups. 
We experimentally demonstrate this for three cases,
each with the decoherence parameter
$p(t) = [1-\exp{(-3t)}]/2$:
\begin{itemize}
    \item Equal mixing with mixing parameters $ x_1=x_2=x_3=0.33$,
    \item unequal mixing with mixing parameters 
$ x_1=x_3=0.3, x_2=0.4$ and 
\item unequal mixing with mixing parameters $ x_1=0.2,  x_2=x_3=0.4$.
\end{itemize}
The $V$ matrix in this case is evaluated to be

\begin{eqnarray}
V = \left(
\begin{array}{cccc}
\sqrt{1-p(t)} & \sqrt{p(t)} & 0 & 0 \\
\sqrt{x_1 p(t)} & -\sqrt{x_1(1-p(t))} & \sqrt{1-x_1} & 0 \\
\sqrt{x_2 p(t)} & -\sqrt{x_2(1-p(t))} & -\sqrt{\frac{x_1 x_2}{1-x_1}} & \sqrt{\frac{x_3}{1-x_1}} \\
\sqrt{x_3 p(t)} & -\sqrt{x_3(1-p(t))} & -\sqrt{\frac{x_1 x_3}{1-x_1}} & -\frac{x_2}{\sqrt{x_2 (1-x_1)}} \\
\end{array}
\right).
\end{eqnarray}
The decay rate of the decoherence parameter $p(t)$ is dependent on the chosen
constant $c$. Therefore, determining the optimal time interval required to study
the behavior of the system  is directly linked to the selection of
$c$. Shorter time periods are preferable to minimize decoherence during
experimental duration. The appropriate choice of $c$ is crucial to effectively
study the impact of the resulting dynamical map on the system, while minimizing noise
interference.

\begin{figure}[H] 
\centering
\includegraphics[scale=1]{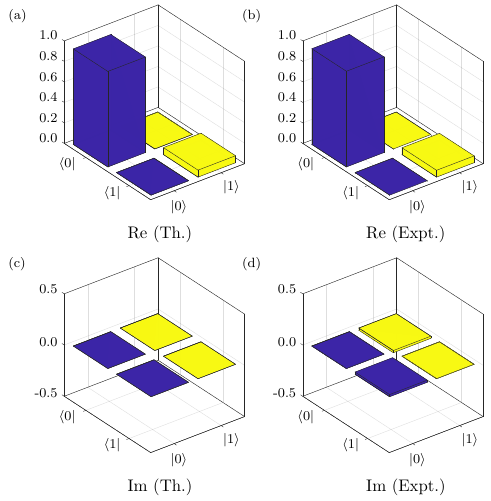}
\caption{Bar plots illustrating the real (Re) and imaginary (Im) components of the theoretical (Th.) and experimental (Expt.) density matrices for the specific case of unequal mixing of two Pauli semigroups at $t=0.1ms$.
} 
\label{tomo}
\end{figure}

The final three-qubit density matrix was reconstructed using the least squares constrained convex optimization method. The average fidelity of the experimental matrices obtained is 0.98 $\pm$ 0.01. The experimental output matrix for the single-qubit
the system is obtained after tracing over the ancilla qubits. We plot bar graphs, Fig.~\ref{tomo} to visually compare the real and imaginary parts of the theoretical and experimental density matrices for the specific example of the second case of mixing two semigroups at $t=0.1 ms$. The fidelity of the experimental state, in this case, is 0.99. The decoherence
parameter $p(t)$ is computed at every time point from the output matrix and the
experimental data is fitted to obtain the experimental parameter $p_e(t)$ and
its time evolution $\dot{p}_e(t)$. The experimental decay rates are subsequently computed with the help of
Eq.~(\ref{ratesdecaythree}).

\begin{figure}[H] 
\centering
\includegraphics[scale=1]{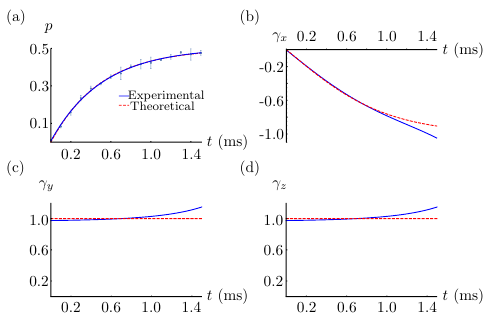}
\caption{Convex Combination of two Pauli semigroups for the case of equal mixing.
(a) Comparison of the theoretical and experimental decoherence parameters $p(t)$. (b) Comparison of theoretical and experimental decay rates $\gamma_1 (t),\gamma_2(t),\gamma_3(t)$ with
mixing parameter $a=0.5$. The red dashed and blue curves represent the
theoretical and the fit to the experimental parameters, respectively.
Experimental data points with error bars are represented by blue dots. The decay rate $\gamma_1(t)$ is negative throughout indicating non-Markovianity.} 
\label{figure2}
\end{figure}

\begin{figure}[H] 
\centering
\includegraphics[scale=1]{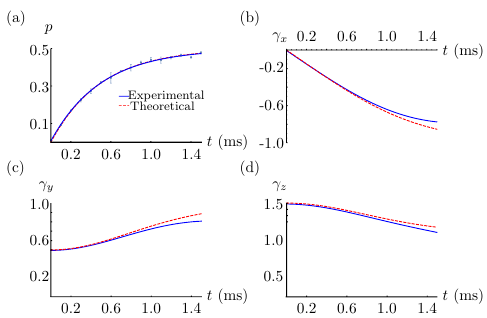}
\caption{Convex Combination of two Pauli dynamical maps for the case of inequal
mixing.  (a) Comparison of the theoretical and experimental decoherence
parameters $p(t)$. (b) Comparison of theoretical and experimental
decay rates $\gamma_1 (t),\gamma_2(t),\gamma_3(t)$ with mixing parameter $a=0.25$. The red dashed and blue curves
represent the theoretical and the fit to the experimental parameters,
respectively.  Experimental data points with error bars are represented by blue
dots. The decay rate $\gamma_1(t)$ is negative throughout indicating non-Markovianity.}
\label{figure3}
\end{figure}

Figures~\ref{figure2} and \ref{figure3} depict a comparison of the theoretical
and experimental results for the two-way mixing case, for equal and
unequal mixing, respectively. For each case, the decoherence parameter $p(t)$ is
plotted in the top panel. The blue dots represent the experimental data with
error bars, the blue curves represent the experimental fits, and the red
dashed curves represent the theoretical parameters. The decay
rates obtained from the experimental data, $\gamma_1(t)$ are negative for both case (i) and case (ii), indicating that the
resultant dynamical map, when two Pauli semigroups maps are mixed, is non-Markovian which
is consistent with the Theorem~1 in ~\cite{jagadish_convex_2020}.

\begin{figure}[H] 
\centering
\includegraphics[scale=1]{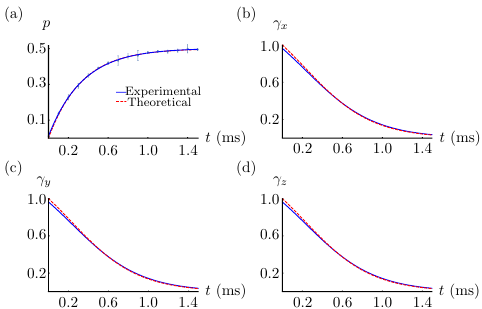}
\caption{Convex Combination of three Pauli dynamical maps for the case of equal
mixing.  (a) Comparison of the theoretical and experimental decoherence
parameters $p(t)$. (b) Comparison of theoretical and experimental
decay rates $\gamma_1 (t),\gamma_2(t),\gamma_3(t)$ with mixing parameters
$x_1=x_2=x_3=0.33$.  The red dashed and blue curves represent the theoretical
and the fit to the experimental parameters, respectively.  Experimental data
points with error bars are represented by blue dots. All the decay rates are positive indicating that the resulting map is Markovian.  }
\label{figure4}
\end{figure}

\begin{figure}[H] 
\centering
\includegraphics[scale=1]{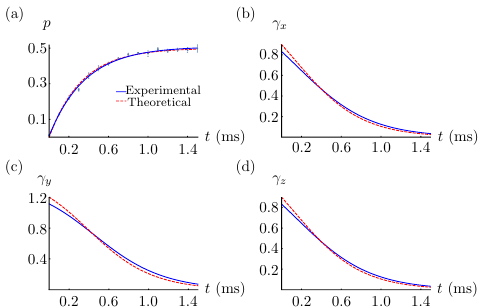}
\caption{Convex Combination of three Pauli dynamical maps for the case of inequal
mixing.  (a) Comparison of the theoretical and experimental decoherence
parameters $p(t)$. (b) Comparison of theoretical and experimental
decay rates $\gamma_1 (t),\gamma_2(t),\gamma_3(t)$ with mixing parameters $x_1=x_3=0.3,
x_2=0.4$, respectively.  The red dashed and blue curves represent the
theoretical and the fit to the experimental parameters, respectively.
Experimental data points with error bars are represented by blue dots. All the decay rates are positive indicating that the resulting map is Markovian. }
\label{figure5}
\end{figure}

\begin{figure}[H] 
\centering
\includegraphics[scale=1]{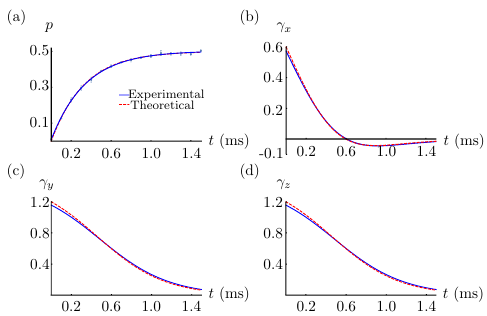}
\caption{Convex Combination of three Pauli dynamical maps for the case of inequal
mixing.  (a) Comparison of the theoretical and experimental decoherence
parameters $p(t)$. (b),(c),(d) Comparison of theoretical and
experimental decay rates $\gamma_1 (t),\gamma_2(t),\gamma_3(t)$ with mixing parameters
$x_1=0.2,x_2=x_3=0.4$, respectively.  The red dashed and blue curves represent
the theoretical and the fit to the experimental parameters, respectively.
Experimental data points with error bars are represented by blue dots.  The
negativity of the decay rate $\gamma_1(t)$ indicates non-Markovianity of the resulting map.} 
\label{figure6}
\end{figure}

Figures~\ref{figure4}-\ref{figure6} presents a comparison of the theoretical and
experimental results for the case of three-way mixing. For each case, the
decoherence parameter $p(t)$ is plotted in the top panel. The blue dots
represent the experimental data with error bars, the blue curves represent the
experimental fits, and the red dashed curves represent the theoretical
parameters. To determine whether the resultant dynamical map is Markovian or
Non-Markovian, the decay rates are analyzed. The decay rates
$\gamma_1 (t),\gamma_2(t),\gamma_3(t)$ were all positive for case (i) and case
(ii) as shown in plots (b),(c) and (d) respectively, indicating that the
resultant dynamical maps are Markovian. However, for case (iii), the negative decay
rate of $\gamma_1 (t)$ suggests that the resultant dynamical map is non-Markovian which is
consistent with the theoretical results.

Figures~\ref{figure2}-\ref{figure6} provide clear evidence of the
agreement between the theoretical and experimental results. The experimental
results clearly corroborate the Markovian or non-Markovian nature of the
dynamical map in both cases of two- and the three-way mixing, which is consistent
with Theorem 1 and the Pauli simplex in~\cite{jagadish_convex_2020}. The
outcomes presented here, which successfully demonstrate the effects of combining
different Pauli semigroups with arbitrary mixing parameters, provide
valuable insights for the study of memory effects in open quantum systems.
Moreover, these results are significant for the development of quantum error
correction and fault-tolerant quantum computing.

\section{Conclusions}
\label{concl}
In our experimental study, we have successfully demonstrated the combination of
two and three Pauli semigroups, with different mixing parameters. The main
objective was to investigate the Markovianity and non-Markovianity of the
resulting dynamical maps. By analyzing the decay rates associated with these
dynamical maps, we were able to assess the characteristics of the quantum maps
under investigation.  We compared our  experimental analysis with the
theoretical predictions. The comparative analysis allowed us to validate the
accuracy of our experimental findings and establish the reliability of our
approach. The good agreement between the experimental results and theoretical
expectations highlight the efficacy of our methodology in capturing the
underlying dynamics of the system-environment interactions.  This research
represents a significant step forward in advancing our understanding of quantum
correlations and the interplay between the system and its surrounding
environment.  Overall, our experimental investigation contributes to the growing
body of knowledge in the field of quantum dynamics, paving the way for further
studies on the characterization and manipulation of quantum information in
realistic environments. NMR, with its precise control, long coherence times and
accurate measurements serves as a good platform for simulating the dynamics of
open quantum systems and understanding the correlations between quantum systems
and their environment.

\acknowledgements
V.J. acknowledges financial support by the Foundation for Polish Science
through TEAM-NET project (contract no. POIR.04.04.00-00-17C1/18-00). R.S. and K.D. acknowledge financial support from Department of Science and Technology (DST), India, Grants Nos:DST/ICPS/QuST/Theme-1/2019/14 and DST/ICPS/QuST/Theme-2/2019/Q-74, respectively. RS also acknowledges the support of the Govt. of India DST/SERB grant CRG/2022/008345.
\vspace{-3 mm}

%
\end{document}